%% file: main.tex
\documentclass[journal]{IEEEtran}
\usepackage[printonlyused]{acronym}

\usepackage{cite}
\usepackage{balance}
\usepackage{algorithmic}
\usepackage{graphicx}
\usepackage{textcomp}
\usepackage{xcolor}
\def\BibTeX{{\rm B\kern-.05em{\sc i\kern-.025em b}\kern-.08em
    T\kern-.1667em\lower.7ex\hbox{E}\kern-.125emX}}

\usepackage{url}
\usepackage{hyperref}
\usepackage{gensymb}
\usepackage{soul} 
\usepackage{booktabs}  
\usepackage{pbox}
\usepackage{subfigure}
\usepackage{epsfig,endnotes}
\usepackage{epstopdf}
\usepackage{times}
\usepackage{color}
\usepackage{colortbl} 
\usepackage{graphicx}
\usepackage{xspace}
\usepackage{enumitem}
\usepackage{multirow}
\usepackage{color}
\usepackage{tikz}
\usepackage[ruled,vlined,linesnumbered]{algorithm2e}
\usepackage{lmodern}
\usepackage[T1]{fontenc}
\usepackage[utf8]{inputenc}
\colorlet{shadecolor}{gray!10}
\colorlet{high}{blue!10}
\colorlet{low}{red!10}
\usepackage{colortbl}
\usepackage{booktabs} 
\usepackage[many]{tcolorbox}
\usepackage{paralist}
\usepackage{tabularx}
\definecolor{orange}{rgb}{1,0.8,0}
\definecolor{darkgreen}{rgb}{0,0.6,0.3}
\usepackage{pifont}
\definecolor{orange}{rgb}{1,0.8,0}
\definecolor{darkgreen}{rgb}{0,0.6,0.3}
\usepackage{subcaption}

\input{sections/acronyms}

\usepackage{siunitx}
\DeclareSIUnit\bps{\bit\per\second}
\DeclareSIUnit\kbps{\kilo\bit\per\second}
\DeclareSIUnit\Mbps{\mega\bit\per\second}
\DeclareSIUnit\Bps{\byte\per\second}
\DeclareSIUnit\kBps{\kilo\byte\per\second}
\DeclareSIUnit\MBps{\mega\byte\per\second}
\DeclareSIUnit\kiBps{\kibi\byte\per\second}
\DeclareSIUnit\MiBps{\mebi\byte\per\second}

\newif\ifcomment
\commenttrue

\ifcomment
    \newcounter{AELNumberOfComments}
    \stepcounter{AELNumberOfComments}
    \newcommand{\ael}[1]{\textcolor{purple}{\small \bf [ael\#\arabic{AELNumberOfComments}\stepcounter{AELNumberOfComments}: #1]}}
    \newcounter{DPNOTENumberOfComments}
    \stepcounter{DPNOTENumberOfComments}
     \newcommand{\dpnote}[1]{\textcolor{orange}{\small \bf [dp\#\arabic{DPNOTENumberOfComments}\stepcounter{DPNOTENumberOfComments}: #1]}}
     
    \newcounter{JOINumberOfComments}
    \stepcounter{JOINumberOfComments}

    \newcounter{JSNumberOfComments}
    \stepcounter{JSNumberOfComments}
    
    \newcommand{\jesus}[1]{\textcolor{blue}{\small \bf [JS\#\arabic{JSNumberOfComments}\stepcounter{JSNumberOfComments}: #1]}}

    \newcounter{SGENumberOfComments}
    \stepcounter{SGENumberOfComments}
    \newcommand{\sge}[1]{\textcolor{cyan}{\small \bf [sge\#\arabic{SGENumberOfComments}\stepcounter{SGENumberOfComments}: #1]}}
    
\else
    \newcommand\dpnote[1]{}
    \newcommand\ael[1]{}
    \newcommand\mc[1]{}
    \newcommand\jesus[1]{}
    \newcommand\sge[1]{}
\fi

\newcommand*\circled[1]{\tikz[baseline=(char.base)]{
            \node[shape=circle,draw,inner sep=1pt] (char) {#1};}}
            
\begin{document}
\title{Anomaly Detection for IoT Global Connectivity}

\author{
    \IEEEauthorblockN{
    \noindent\makebox[\textwidth][c]{%
        \begin{minipage}{0.8\linewidth}
        \centering
            Jesus Omaña Iglesias\IEEEauthorrefmark{1},
            Carlos Segura Perales\IEEEauthorrefmark{1},
            Stefan Gei\ss{}ler\IEEEauthorrefmark{2},
            Diego Perino\IEEEauthorrefmark{1},\\
            Andra Lutu\IEEEauthorrefmark{1}
        \end{minipage}
    }
    }
    \newline
    \IEEEauthorblockA{
        \IEEEauthorrefmark{1}Telefonica, Spain\\
    }
    \IEEEauthorblockA{
        \IEEEauthorrefmark{2}University of W\"urzburg, Germany\\
    }
}

\maketitle

\begin{abstract}
    Internet of Things (IoT) application providers rely on Mobile Network Operators (MNOs) and roaming infrastructures to deliver their services globally. In this complex ecosystem, where the end-to-end communication path traverses multiple entities, it became increasingly challenging to guarantee communication availability and reliability. 
    Further, most platform operators use a \textit{reactive} approach to communication issues, responding to user complaints only after incidents have become severe, compromising service quality. 
    This paper presents our experience in the design and deployment of ANCHOR -- an \textit{unsupervised} anomaly detection solution for the IoT connectivity service of a large global roaming platform. 
    ANCHOR assists engineers by filtering vast amounts of data to identify potential problematic clients (i.e., those with connectivity issues affecting several of their IoT devices), enabling proactive issue resolution before the service is critically impacted. We first describe the IoT service, infrastructure, and network visibility of the IoT connectivity provider we operate. 
    Second, we describe the main challenges and operational requirements for designing an unsupervised anomaly detection solution on this platform. Following these guidelines, we propose different statistical rules, and machine- and deep-learning models for IoT verticals anomaly detection based on passive signaling traffic. We describe the steps we followed working with the operational teams on the design and evaluation of our solution on the operational platform, and report an evaluation on operational IoT customers.
    \end{abstract}

\maketitle

\input{sections/01_introduction}

\input{sections/02_background}

\input{sections/03_requirements}
\input{sections/04_model_design}

\input{sections/05_ml_evaluation2}
\input{sections/06_discussion}

\input{sections/07_related}

\input{sections/08_conclusions}


\small
\balance
\bibliographystyle{acm}

\input{main.bbl}


\end{document}

%% file: sections/acronyms.tex
\acrodef{2G}{Second Generation}
\acrodef{3G}{Third Generation}
\acrodef{4G}{Fourth Generation}
\acrodef{5G}{Firth Generation}
\acrodef{ADB}{Android Debug Bridge}
\acrodef{APN}{Access Point Name}
\acrodef{API}{Application Programming Interface}
\acrodef{ASCI}{Advertising Standards Council of India}
\acrodef{ASN}{Autonomous System Number}
\acrodef{ASP}{Application Service Providers}
\acrodef{CDF}{Cumulative Distribution Function}
\acrodef{CDN}{Content Delivery Network}
\acrodef{CDR}{Call Detail Record}
\acrodef{CL}{Cancel Location}
\acrodef{CN}{Core Network}
\acrodef{CNN}{Convolutional Neural Network}
\acrodef{DCH}{Data Clearing House}
\acrodef{DL}{Downlink}
\acrodef{DNS}{Domain Name Service}
\acrodef{E2E}{End-to-end}
\acrodef{ECDF}{Empirical Cumulative Distribution Function}
\acrodef{EC}{European Commission}
\acrodef{EMSISDN}{Encrypted Mobile Station International Subscriber Directory Number}
\acrodef{EU}{European Union}
\acrodef{EaaS}{Experiment as a Service}
\acrodef{FNO}{Fixed Network Operator}
\acrodef{FQDN}{Fully Qualified Domain Name}
\acrodef{GGSN}{Gateway GPRS Support Node}
\acrodef{GMM}{Gaussian Mixture Model}
\acrodef{GPRS}{General Packet Radio Service}
\acrodef{GRX}{\ac{GPRS} Roaming eXchange}
\acrodef{GSMA}{\ac{GSM} Association}
\acrodef{GSM}{Global System for Mobile communications}
\acrodef{GTP}{GPRS Tunneling Protocol}
\acrodef{HLR}{Home Location Registry}
\acrodef{HMNO}{Home Mobile Network Operator}
\acrodef{HR}{Home-routed Roaming}
\acrodef{HTTP}{Hyper Text Transfer Protocol}
\acrodef{ICMP}{Internet Control Message Protocol}
\acrodef{IHBO}{IPX hub breakout}
\acrodef{IMEI}{International Mobile Equipment Identity}
\acrodef{IMSI}{International Mobile Subscriber Identity}
\acrodef{IOT}{Inter Operator Tariff}
\acrodef{IPG}{Inter-Packet Gap}
\acrodef{IPX-P}{\ac{IPX} Provider}
\acrodef{IPX}{IP Packet Exchange}
\acrodef{IRB}{Institutional Review Board}
\acrodef{ISP}{Internet Service Providers}
\acrodef{ISP}{Internet Service Provider}
\acrodef{IXP}{Internet Exchange Point}
\acrodef{IoT}{Internet of Things}
\acrodef{KS}{Kolmogorov-Smirnov}
\acrodef{LBO}{local breakout}
\acrodef{LPWA}{Low Power Wide Area}
\acrodef{LTE-M}{\ac{LTE} Machine Type Communication}
\acrodef{LTE}{Long Term Evolution}
\acrodef{LSTM}{Long Short-Term Memory}
\acrodef{M2M}{Machine-to-Machine}
\acrodef{MAP}{Mobile Application Part}
\acrodef{MBB}{Mobile Broadband}
\acrodef{MCCMNC}{\ac{MCC}-\ac{MNC}}
\acrodef{MCC}{Mobile Country Code}
\acrodef{MME}{Mobility Management Entity}
\acrodef{MNC}{Mobile Network Code}
\acrodef{MNO}{Mobile Network Operator}
\acrodef{MSC}{Mobile-services Switching Centre}
\acrodef{MSISDN}{Mobile Station International Subscriber Directory Number}
\acrodef{MTC}{Machine Type Communications}
\acrodef{MVNO}{Mobile Virtual Network Operator}
\acrodef{NB-IoT}{Narrow Band \ac{IoT}}
\acrodef{NDT}{Network Diagnostic Tool}
\acrodef{NRA}{National Regulatory Authority}
\acrodef{OS}{Operating System}
\acrodef{PCA}{Principal Component Analysis}
\acrodef{PDP}{Packet Data Protocol}
\acrodef{PGW}{Packet Data Network Gateway}
\acrodef{PoP}{Point of Presence}
\acrodef{QCI}{QoS Class Identifier}
\acrodef{QoE}{Quality of Experience}
\acrodef{QoS}{Quality of Service}
\acrodef{RAN}{Radio Access Network}
\acrodef{RAT}{Radio Access Technology}
\acrodef{RMBT}{RTR Multithreaded Broadband Test}
\acrodef{RTT}{Round-Trip Time}
\acrodef{SAI}{Send Authentication Information}
\acrodef{SAIR}{Send Authentication Information Request}
\acrodef{SCCP}{Signaling Connection Control Part}
\acrodef{SGSN}{Serving GPRS Support Node}
\acrodef{SGW}{Serving Gateway}
\acrodef{SIMs}{Subscriber Identity Modules}
\acrodef{SIM}{Subscriber Identity Module}
\acrodef{SIP}{Session Initiation Protocol}
\acrodef{SLA}{Service-Level Agreement}
\acrodef{TAC}{Type Allocation Code}
\acrodef{TAP}{Transferred Account Procedure}
\acrodef{TCP}{Transmission Control Protocol}
\acrodef{UDP}{User Datagram Protocol}
\acrodef{UE}{User Equipment}
\acrodef{UL-GPRS}[UL\_GPRS]{Update \ac{GPRS} Location}
\acrodef{UL}{Update Location}
\acrodef{VAE}{Variational Autoencoder}
\acrodef{VLR}{Visited Location Registry}
\acrodef{VMNO}{Visited Mobile Network Operator}
\acrodef{VoIP}{Voice over IP}
\acrodef{VoLTE}{Voice over \ac{LTE}}
\acrodef{xDR}{eXtended Detail Record}

%% file: sections/01_introduction.tex

\section{Introduction}


In the current cellular ecosystem, mobile networks jointly provide smart connectivity to a dazzling number of heterogeneous devices that operate globally, in different environments, and with varying performance requirements. 
One of the main drivers behind this need for global, ubiquitous cellular connectivity comes from the success of the \ac{IoT}, which the community correctly predicted ever since 2015. 
However, one aspect that the 5G architecture did not foresee is the prevalent \textit{global} deployment model that IoT vertical applications have been demanding, which depends on international roaming. 
IoT connectivity providers cater to the IoT vertical applications demand for global reach (i.e., the ability to deploy their devices in multiple economies across the world) and low management overhead by overloading the international roaming function, which is in-built within the cellular ecosystem. 
Roaming allows for a global deployment approach because it enables devices to connect to any \ac{RAN} provider, while depending on the same home operator (i.e., the IoT provider). 
As a result, the once niche service of international roaming now sustains the growth of the gargantuan IoT market~\cite{lutu2020things, andrade2017connected, kolamunna2018first}, and powers attractive products for the IoT vertical applications, such as the Global IoT SIM~\cite{alcala22global}. 

Though this model provides the global reach and simplified management that IoT verticals desire, it also introduces significant complexity in detecting and addressing anomalies. This complexity arises partly because end-to-end data paths, which were once contained within a single operator’s domain, now span multiple domains, and depend on resources from various entities -- specifically, the visited \ac{RAN} provider, the IoT provider's home core network operator, and the international roaming hub that interconnects the two before-mentioned roaming partners -- requiring their coordination to enable the IoT device connectivity.
These globally distributed, interconnected and independently operated systems are prone to performance degradation and outages, even when applying best practices for their operation.
Hence, we highlight here the stringent need for high-performing anomaly detection tools to support the operations of the IoT provider~\cite{Vomhoff22NOMS}, and go beyond the functionality offered by existing alarm systems.  

\begin{figure}[!t]
    \centering
    \includegraphics[width=\linewidth]{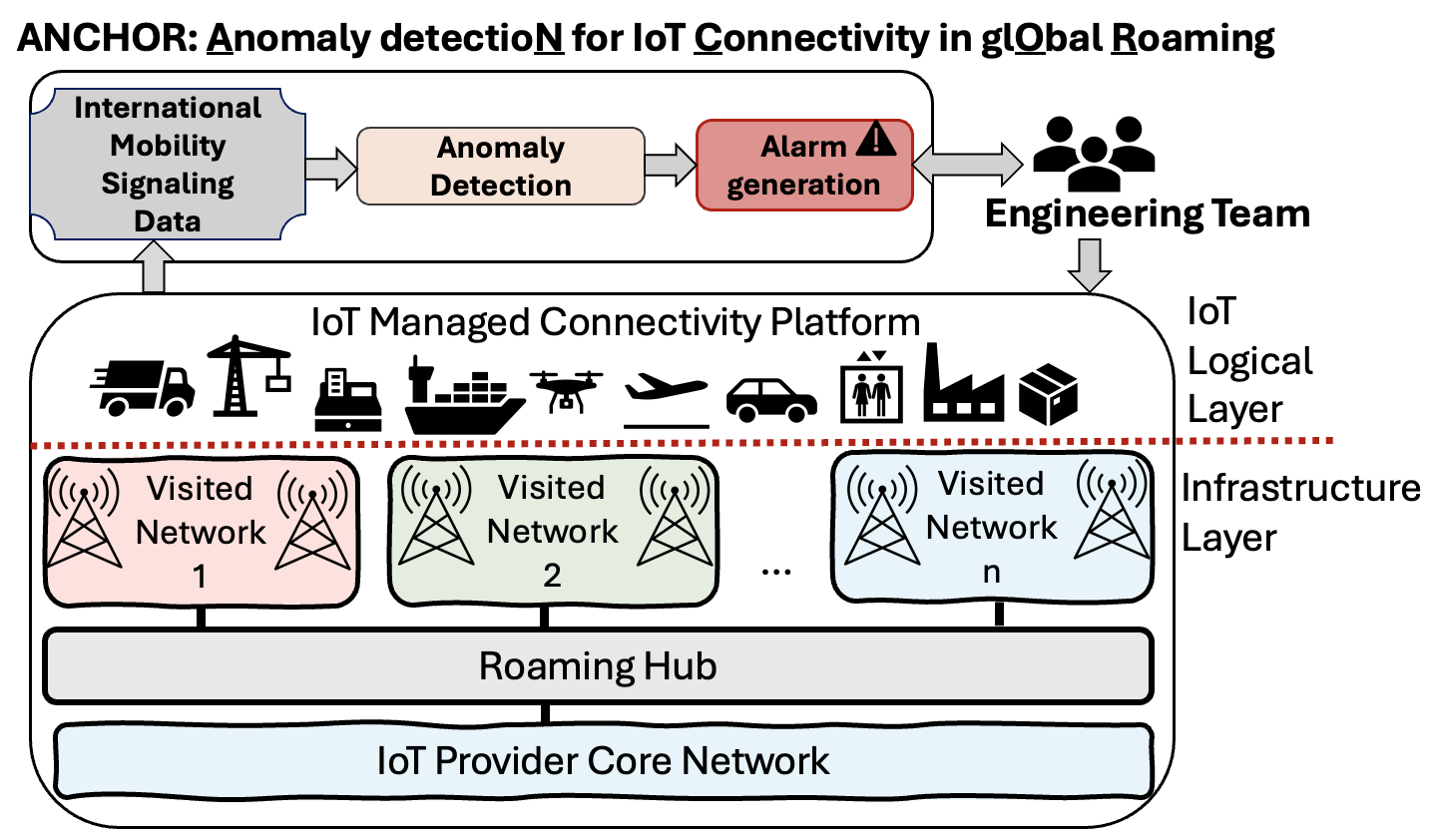}
        \vspace{-6mm}
    \caption{\small ANCHOR for IoT applications (e.g., fleet tracking, cargo ships, drones, planes, connected cars). }
	\label{fig:anchor_overview}
    \vspace{-3mm}
\end{figure}

In this paper, we build an \textit{unsupervised} anomaly detection pipeline that relies on machine learning approaches to improve the managed connectivity service of a global \ac{IoT} service provider (see Figure~\ref{fig:anchor_overview}). 
Our solution for \textit{\ul{a}nomaly detectio\ul{n} for IoT \ul{c}onnectivity in gl\ul{o}bal \ul{r}oaming (ANCHOR)} aims to detect issues within the ecosystem before they become critical (i.e., before the IoT client suffers permanent service outages).
We collect international mobility signaling data from the \textit{roaming hub} that links the IoT provider to its roaming partners, using this interconnection point to gather control plane insights on the connectivity and mobility of IoT devices~\cite{geissler21signaling}.
Specifically, by collecting signaling traffic for international mobility and authentication from different IoT devices (e.g., smart sensors, connected cars, shipping containers), we build an anomaly detection approach to generate alarms that the engineering team operating the IoT platform can use to further improve their service.

We contribute our experience with designing, implementing, and evaluating this pipeline together with the engineering team of the telco \ac{IoT} operator.


\textbf{Overview of ANCHOR}. 
We propose an unsupervised anomaly detection method, and share our experience with building the solution with various data representations that leverage the strengths of different models. 
ANCHOR captures anomalies at the IoT client level, considering the behavior of their entire fleet of IoT devices active in a specific economy. 
Given the diversity of IoT devices (e.g., elevators have distinct mobility patterns compared to vehicles), we employ a two-step approach for ANCHOR: first, we identify the context of an observation (and associated baseline behavior), and then assign its anomaly status (based on the deviation from baseline). Our results demonstrate the effectiveness of expert knowledge in feature engineering, enabling simpler machine learning models, such as isolation forests, to detect anomalies in four out of five IoT clients —all cases that were missed by the network operator's alarm systems.

We provide next a short overview of the IoT provider platform architecture and present our data collection approach (Section~\ref{sec:background}).
We rely on a comprehensive, unique dataset that records the signaling traffic of a variety of IoT devices operating worldwide (spanning 40 countries) for multiple IoT vertical applications (Section~\ref{sec:dataset}). 
To meet the requirements of the engineering team (Section~\ref{sec:requirements}), we compare as part of the ANCHOR framework two different approaches for data representation and different models (Section~\ref{sec:model_design}). 
Given the heterogeneity of IoT devices, we integrate in each approach a clustering algorithm to automatically detect the context of the devices, based on mobility patterns and message exchange volumes. This allows us to identify emerging patterns, and thus design an anomaly detection approach that is application-agnostic.
We show that the ANCHOR pipeline enables platform operators to avoid incidences before they become critical, and to learn about new anomalies not detected by alarm systems (Section~\ref{sec:ml_results}).
ANCHOR offers a proactive approach for managed IoT connectivity services, which we tested and integrated in the analytics platform operated by the IoT provider (Section~\ref{sec:discussion}).


%% file: sections/02_background.tex

\vspace{-2mm}
\section{Data Collection}
\label{sec:background}

In this section, we provide an overview of the ecosystem structure that enables our IoT provider to deliver global connectivity to millions of IoT devices worldwide. We discuss the various anomalies that can occur in the context of global IoT connectivity, and describe the ground truth dataset we have gathered.
We introduce the monitoring dataset collected from the IoT provider, designed to capture symptoms of anomalies. 

\begin{figure}[!t]
    \centering
    \includegraphics[width=\linewidth]{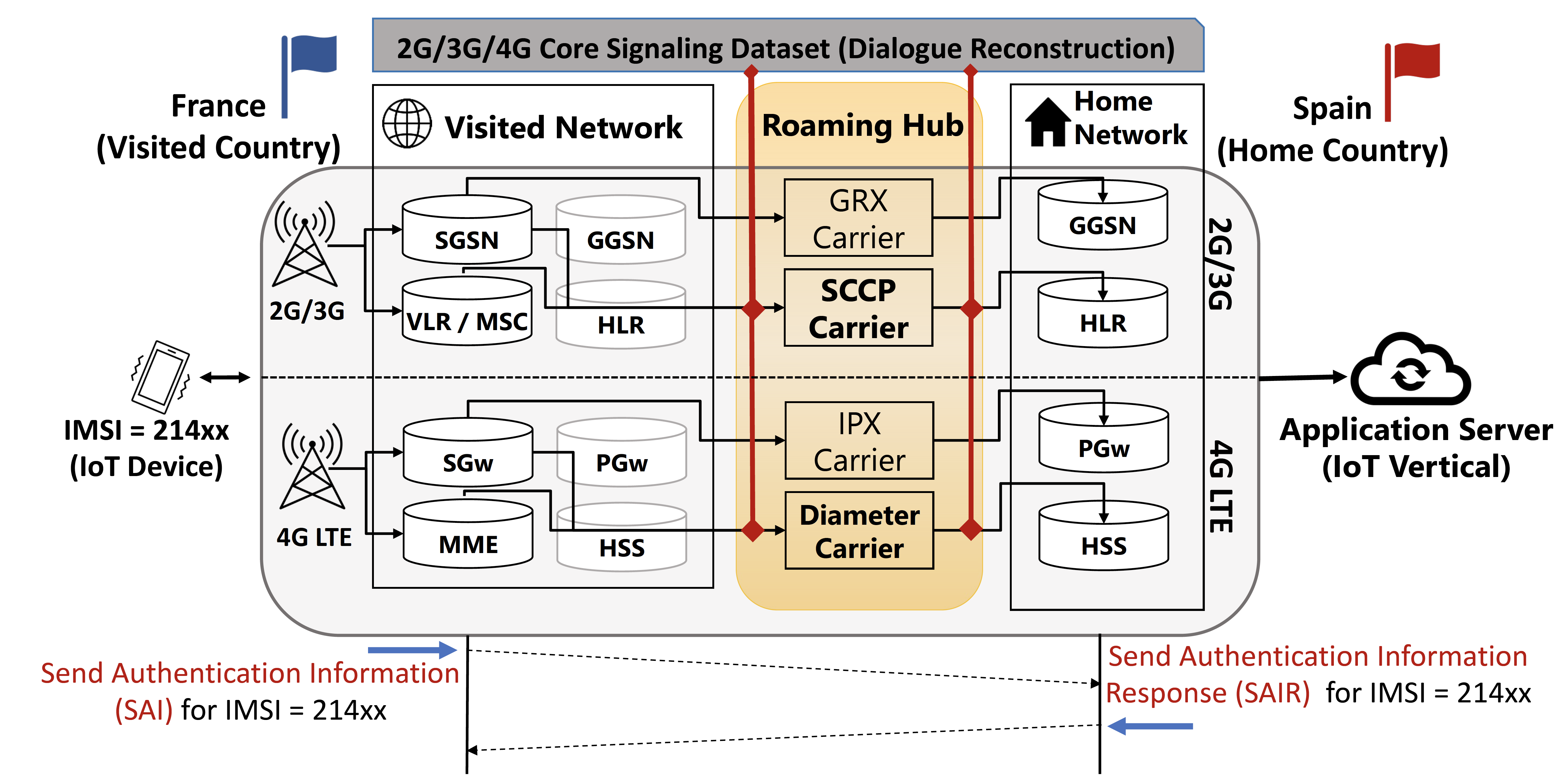}
        \vspace{-7mm}
    \caption{\small Overview of the IoT cellular ecosystem and the role of roaming for global IoT connectivity. We show the interfaces we monitor to build the dataset of signaling traffic (\ac{SCCP} and Diameter Carriers for 2G/3G and 4G signaling, respectively). In the lower part, we show an example of 2G/3G signaling \textit{dialogue} between the deployment location in France (the visited network) and the core network location in Spain (the home network) to authenticate the IoT device whose \ac{SIM} has IMSI 214xx. }
	\label{fig:iot_ecosystem}
\end{figure}

\vspace{-3mm}
\subsection{IoT Managed Connectivity}
\label{sec:iot_platforms}

Support for ``things'' operating globally has become critical for \ac{IoT} verticals, from connected cars to smart sensors to wearables~\cite{lutu2020things, kolamunna2018first, andrade2017connected}.
IoT verticals require world-wide deployment, while keeping operational simplicity in terms of managing the connectivity of their devices and customers.
IoT managed connectivity providers (or, in short, \textit{IoT providers}, such as Twilio \cite{twilio} or EMnify\cite{emnify}) answer to these requirements by exploiting the cellular ecosystem's roaming function~\cite{lutu2020things, lutu21insights}.
As a result, the communications between IoT devices and applications depend on multiple network domains, often operated by different organizations\cite{lutu2020ipx, vomhoff24shortcut}. 
In this ecosystem, we distinguish between the \textit{IoT logical layer}, which includes IoT applications and their connectivity needs, and the \textit{infrastructure layer}, which supports the IoT provider services (see (Fig~\ref{fig:anchor_overview}).

On the \emph{IoT logical layer}, IoT customers (e.g., car manufacturers, fleet tracking providers, connected elevator manufacturers) contract managed connectivity from IoT providers ~\cite{kite,emnify,hologram}. These platforms provide their customers connectivity via a set of \acp{SIM} that can be installed in IoT devices and used globally (also called global \acp{SIM}). 
They also provide management in a centralized manner and other added values services (e.g., security, APIs) for the \acp{SIM} provisioned to every customer.

IoT providers do not own the entire \emph{infrastructure layer}, but contract global \acp{SIM} from one or multiple \acp{MNO}, which operate the IoT Provider Core Network. These \acp{MNO} provide the SIMs' unique identifier, the \ac{IMSI}.
The MNOs are therefore the \textit{home networks} enabling connectivity for IoT devices operating world-wide~\cite{lutu2020things}, as we show in Figure~\ref{fig:anchor_overview}.
Since most IoT verticals deploy devices outside the SIMs' home country, \acp{MNO} use their roaming agreements to guarantee global SIMs connectivity. 
International telco carriers that offer the roaming hub service~\cite{lutu2020ipx} interconnect the home network with any other visited network, thus increasing the geographical footprint of the IoT provides outside the borders of their respective MNOs' home countries (see Figure~\ref{fig:iot_ecosystem}).  
In this paper, we analyze the IoT managed connectivity service of a large operational IoT provider whose service depends on a home network in Europe, and a roaming hub with more than 100 \acp{PoP}\footnote{https://www.broadbandsearch.net/definitions/point-of-presence} in 40+ countries world-wide.

\vspace{-3mm}
\subsection{IoT Connectivity Anomalies}
\label{sec:ground-truth}

The IoT provider's ticketing system tracks the incidents handled by the engineering teams, who monitor the IoT provider service. 
In this paper, we leverage this information to build the ground-truth dataset, which captures anomalies that escape the deployed (threshold-based) guardrails. 

\textit{We define an anomaly as any unplanned incident that partially or fully disrupts the connectivity of an IoT client's fleet of devices within a specific country.}

Though root cause detection is beyond the scope of this work, we group the tickets based on the part of the ecosystem where the issue was rooted. In our case, this can include any entity within Figure~\ref{fig:anchor_overview}, namely the IoT logical layer components (e.g., the IoT provider platform \ac{API}, billing and data quota configurations, etc.), the roaming hub, the visited or home networks. 
The reported incidents associate with data connectivity issues (i.e., IoT devices suffering from bad coverage in the deployment area), roaming-specific issues related to the commercial agreements between the roaming partners (i.e., the home operator and the visited operator), billing issues (e.g., some devices report a much larger amount of traffic than agreed upon in the billing model), or IoT provider \ac{API} issues (e.g., the client cannot access to re-configure its \acp{SIM}), the active \acp{IMSI} were not correctly configured).

\textbf{Critical anomalies.} These incidents often trickle down to impact the affected devices' signaling traffic patterns. For example, in an attempt to regain connectivity and restore the baseline services, an IoT device might generate a signaling traffic storm in the visited radio network. 
The consequences of such issues can be severe, impacting not only the number of affected devices but also leading to potential disruptions. 
For instance, aggressive signaling behavior can endanger the roaming agreements between partner MNOs, which are very expensive and time-consuming to remedy.
Unpredictable outages affect the IoT provider's availability and reliability metrics (thus impacting its \acp{SLA}), and they translate into entire fleets of devices associated to IoT customers malfunctioning. 
Another example is that a misconfiguration in the traffic routing within the IoT provider could result in the emergency phone lines installed in the elevators operated by an IoT client becoming unavailable.

In order to separate the critical anomalies that impact the availability score of the IoT platform, the IoT provider follows several heuristics derived from the domain knowledge provided by the engineering teams. 
The critical anomalies relate to three main parts of the ecosystem (Figure~\ref{fig:iot_ecosystem}): global network issue (i.e., the roaming hub), platform issue (i.e., the IoT Provider commercial service in the IoT logical layer), or issue in the network of an MNO supporting the IoT provider (i.e., the home or the visited network). The critical anomalies include, for example, success rate of \ac{MAP} procedures below 25\% in the \ac{HLR}, or services (voice, data or SMS) impacted for more than 5\% of all the devices provisioned. 
Despite these heuristics being in place, IoT customers (with various device fleet sizes) trigger service tickets when they observe irregularities in the behavior of their devices. During October 2022, we observed more than 90 tickets triggered by IoT customers operating their own IoT vertical applications. Out of these, approximately 14\% were eventually abandoned, either because of lack of information (e.g., the customer does not specify the devices affected by the anomalies reported), or because the anomaly reported was a transient one.  

\input{sections/anomalies_table}

\textbf{Ground Truth.} For validating the ANCHOR pipeline, we build a dataset of ground-truth anomalies that the IoT provider learned about only after a notification from the customer. We build this dataset by analyzing the tickets triggered throughout October 2022, and identifying the affected devices, using their associated \acp{IMSI}, and the duration of the anomaly. 
For this exercise, we focus on five specific IoT customers/clients that serve different final end-users (including, for example, shipment tracking solutions, emergency communication, or smart sensors). 
Table~\ref{tab:anomalies_groundTruth} gives more details about the ground truth dataset. 
Additionally, we selected a \textit{control} client known to have operated smoothly during this period.
We selected anomalies with medium, high, or critical impact that affected from a few tens to a few thousand devices within different countries (including Spain, India, Argentina or El Salvador). 
In this period there were, on average, more than $2.9$ million daily active devices (and more than $4$ million devices overall).

\vspace{-3mm}
\subsection{Monitoring at the Roaming Hub}

To enable ANCHOR, we collect signaling data flowing through the roaming hub to capture the connectivity status of IoT devices belonging to the real-world IoT provider we monitor. 
Specifically, we distinguish the IoT devices of interest from the point of the roaming hub using their unique identifiers (i.e., encrypted \ac{MSISDN} or \ac{IMSI}), and monitor the status of the following core signaling services. 


\noindent \textbf{\ac{SCCP} Global Signaling}. This service provides the signalling capabilities for 2G-3G technologies. The IPX provider monitors the \ac{MAP} protocol and specifically the traffic corresponding to the following procedures:
i) location management (e.g., \ac{UL}, \ac{UL-GPRS}, \ac{CL}, Purge Mobile Device); ii)  authentication and security (e.g., \ac{SAI}); iii) fault recovery (i.e., by monitoring the result codes of \ac{MAP} requests).

\noindent \textbf{Diameter Exchange}. This service provides signalling capabilities for 4G roaming. The IPX-P collect traffic corresponding to events including Session Initiation Protocol (SIP) Registration, Voice over IP (VoIP) Call, Diameter Transaction, Domain Name Service (DNS) Query or Rich Communication Services (RCS) Session.

\begin{figure*}[t]
\centering
\begin{minipage}[t]{0.29\textwidth}
  \centering
  \includegraphics[width=\linewidth]{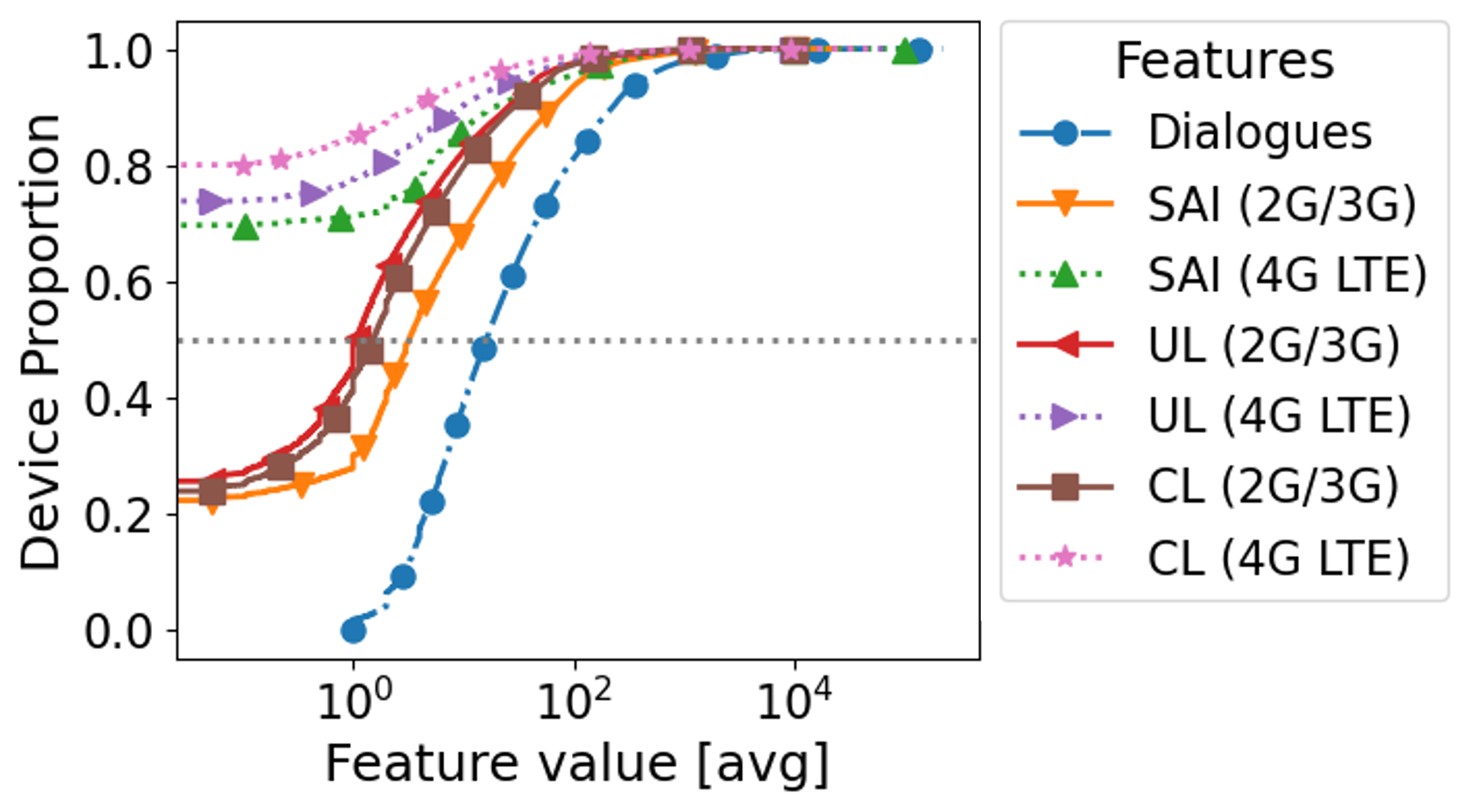}
  \caption*{\small{(a) Empirical CDFs of dialogue volume metrics per device (average number of dialogues per day).}}
  \label{fig:figure1}
\end{minipage}
\hfill
\begin{minipage}[t]{0.34\textwidth}
  \centering
  \includegraphics[width=\linewidth]{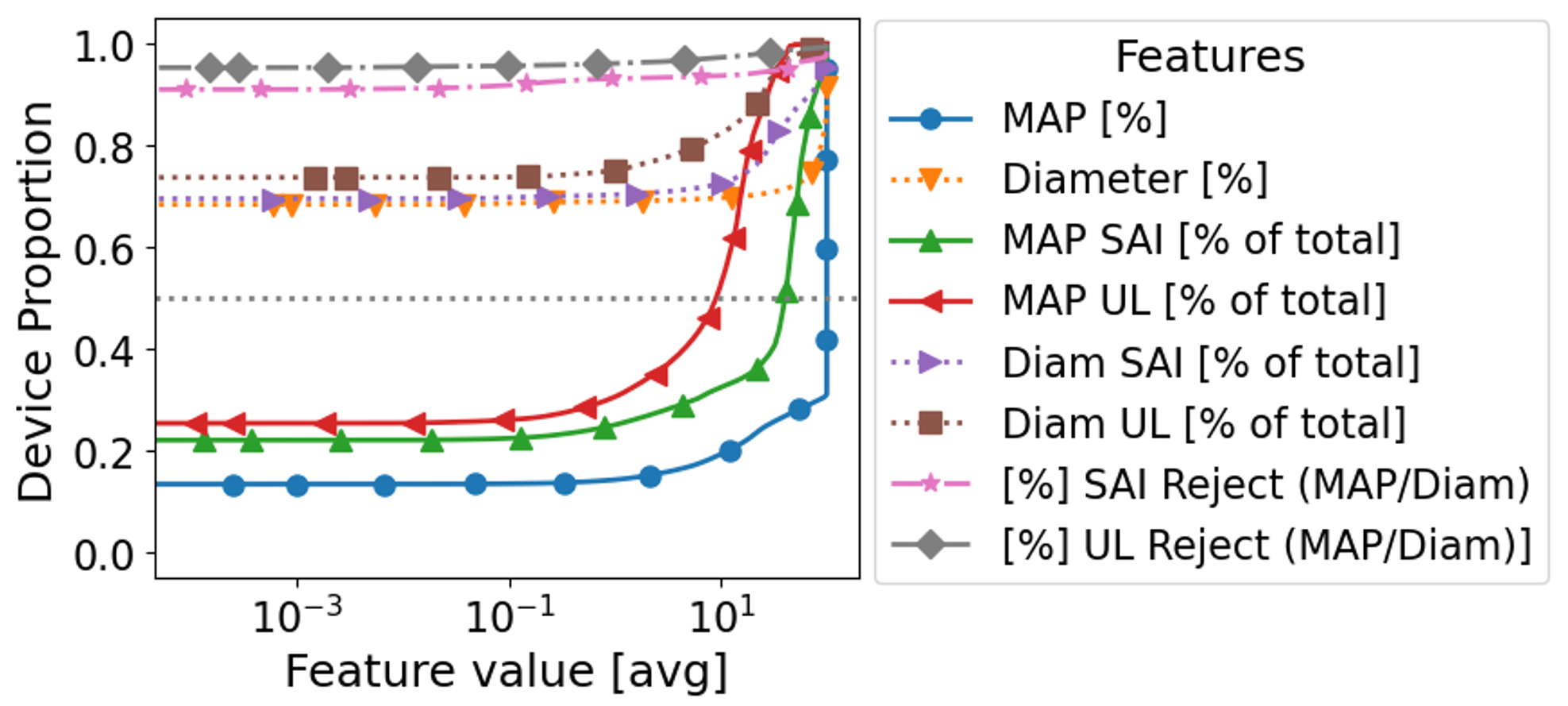}
  \caption*{\small{(b) Empirical CDFs of message type statistics per device. A rejected dialogue is a dialogue ending with a non-successful result.}}
  \label{fig:figure2}
\end{minipage}
\hfill
\begin{minipage}[t]{0.34\textwidth}
  \centering
  \includegraphics[width=\linewidth]{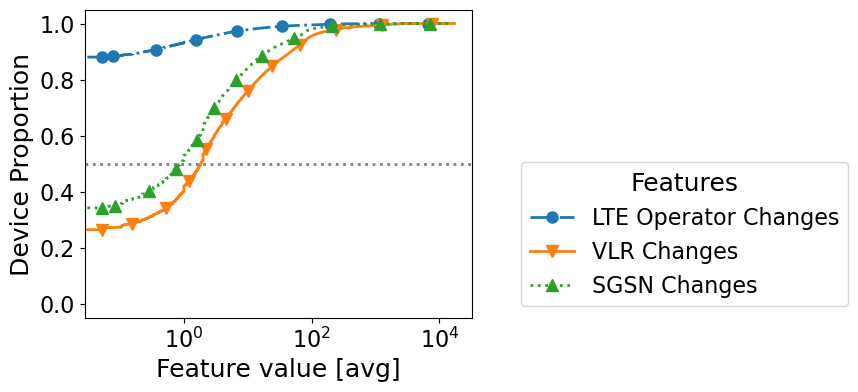}
  \caption*{\small{(c) Empirical CDFs of device mobility metrics.}}
  \label{fig:figure3}
\end{minipage}
\caption{\small Empirical CDFs of selected features from overall statistics and message type statistics.}
\label{fig:features_overall_statistics}
\vspace{-4mm}
\end{figure*}

\vspace{-3mm}
\subsection{Dataset}
\label{sec:dataset}

In this work, we capture from the IoT operator's mobile core network ingress the signaling traffic corresponding to a large sample of \ac{IoT} devices.
We identify each IoT device via its unique identifier (i.e., the encrypted \ac{IMSI})
and capture the traffic corresponding to two signaling services, namely \ac{SCCP} signaling (for 2G/3G connectivity) and Diameter signaling (for 4G LTE connectivity).

We collect our training dataset during September-October 2022. 
It contains signaling data from a sample of $\approx$4M IoT devices from various IoT vertical application that the IoT provider's clients support. 
The dataset comprises over 10 billion signaling interactions, each referred to as a dialogue, which consists of request-response pairs. 

To better understand the scope and methodology of our data collection, Figure~\ref{fig:iot_ecosystem} schematically illustrates the locations and interfaces where we collect our data.
We capture the raw signaling traffic and process it to reconstruct the signaling dialogues between core network functions in the home network and those in the visited network, whose \ac{RAN} is used by IoT devices. 

Considering \ac{SCCP} signaling (for 2G/3G connectivity), we monitor the \ac{MAP} protocol that supports end-user mobility and allows network elements (e.g., the \ac{HLR}, \ac{VLR}, \ac{SGSN}, \ac{MSC}) to communicate. 
We collect traffic corresponding to the procedures of each device:  i) authentication and security (\acl{SAI}), and ii) location management (\acl{UL}, \acl{CL}).
For 4G LTE connectivity, we monitor the Diameter protocol that provides a similar functionality to SCCP. We collect signaling dialogues corresponding to the same procedures we mentioned above for \ac{SCCP}. 

We present in Figure~\ref{fig:features_overall_statistics} the cumulative distribution per device of average monthly dialogue, categorized per dialogue type and corresponding to different generations of radio technologies. 
Figure~\ref{fig:features_overall_statistics}.a illustrates the breakdown of the average number of dialogues per device per day for three signaling message types across \ac{RAT}, namely \ac{SAI}, \ac{UL} and \ac{CL}. 
We observe a clear separation between 2G/3G and 4G LTE dialogue volumes: in our sample, devices are predominantly active on 2G/3G connections. 
In fact, when analyzing the message type statistics in Figure~\ref{fig:features_overall_statistics}.b for MAP and Diameter average percentage of dialogues per device over one month, we observe that 85\% of devices are active in 2G/3G, while only about 30\% of them are active in 4G LTE. 
For capturing device mobility, we use a proxy that tracks changes in the LTE \ac{RAN} provider per device per day, or the changes in the visited 2G/3G network infrastructure (i.e., \ac{VLR} or \ac{SGSN}) the devices use. 
Figure~\ref{fig:features_overall_statistics}.c shows that within our sample, less than 10\% present 4G LTE mobility, while more than 65\% have 2G/3G mobility. Given the dataset's origin from a live production environment, there are certain limitations that must be considered when working with this data.

\textit{Missing Information.} Due to the nature of the dataset's source environments, detailed information about the specific use cases of individual devices is typically unavailable. Operators do not track customers' applications within the aggregated traffic mix. While we may know the identity of the customers (providing clues about the IoT application vertical they serve) and their connectivity requirements (translated to \ac{SLA}), we have no visibility into the actual application of the end-user device. We do not capture data connectivity control plane information (e.g., the \ac{GTP} signaling information) or the user plane information. 

\textit{Geographical Bias.} As a result of investigating a specific IoT provider platform, its customer base exhibits geographical bias that may impact the observed behavior. The traffic we capture for the IoT provider contains mostly devices deployed in more than 30 countries in Europe, Central America and South America.

\vspace{-3mm}
\subsection{Ethical Considerations}
\label{sec:ethical}
Data collection and retention are in accordance with the terms and conditions of the IoT provider and the local regulations, and only with the specific purpose of providing and managing the IoT services. The terms also include data processing for monitoring and reporting as allowed usages of collected data. Data processing only extracts aggregated information, and we do not have access to any personally identifiable information.

%% file: sections/anomalies_table.tex


\newcommand{\No}{\cellcolor{red!25}No\xspace}
\newcommand{\Yes}{\cellcolor{green!25}Yes\xspace}
\newcommand{\YesNo}{\cellcolor{orange!25}Yes*\xspace}
\newcommand{\NA}{\cellcolor[gray]{0.75}N/A\xspace}
\newcommand{\NoEmtpy}{\cellcolor{red!25}\xspace}
\newcommand{\YesEmtpy}{\cellcolor{green!25}\xspace}
\newcommand{\YesNoEmtpy}{\cellcolor{orange!25}\xspace}
\newcommand{\NAEmtpy}{\cellcolor[gray]{0.75}\xspace}

\begin{table*}[t]
\footnotesize
\setlength\tabcolsep{3pt}
\caption{\small Ground truth anomalies based on historical incidents managed in the IoT platform's ticketing system.  }
\label{tab:anomalies_groundTruth}
\vspace{-2mm}
\begin{tabular}{p{1.5cm}| p{2cm} | p{10.5cm} | p{1.3cm} | p{2cm}}
\toprule
    \textbf{Client} & \textbf{Date Reported} & \textbf{Anomaly Description} & \textbf{Devices Location} & \textbf{IoT Vertical} \\ \hline \hline
\toprule
    Client\#1  & 5-6.10.2024 & SIM cards cannot connect to the radio network, nor receive data due to a mis-configured roaming agreement.    & Spain & Smart building \\\hline 
    Client\#2  & 20.10.2022 &  Devices lost connectivity because of transient error in the platform.
   & Argentina & Connected fleet management \\\hline 
    Client\#3  & 27-28.10.2022 & Anomalous Service Reject dialogues from triggered Cancel Location procedure (signaling protocol issues).
  &  India & Smart Building\\\hline  
    Client\#3  & 26.10.2022 & Specific type of traffic blocked (signaling protocol issues).
  &  Germany & Smart Building\\\hline  
    Client\#4  & 24.10.2022 & A few aggressive devices (high signaling volume) because of lack of radio coverage.   &  El Salvador & Smart meters \\\hline 
    Client\#5  & 10.10.2022 & Massive outage due to misconfiguration in one of the server in the IoT Provider platform. Devices were continuously registering at the circuit level, with correct authentication messages, but anomalous Cancel Location from the home network. 
   & USA &  Connected fleet management\\\hline  
\bottomrule
\end{tabular}
\vspace{-3mm}
\end{table*}

%% file: sections/03_requirements.tex
\section{ANCHOR Requirements}
\label{sec:requirements}

In this section, we describe the main requirements and challenges towards building the ANCHOR anomaly detection solution, as well as the approach we take to address them.

\vspace{.5mm}
\noindent \textbf{Vertical-agnostic approach.} 
The ANCHOR solution must be versatile enough to monitor anomalies in various IoT vertical clients. For instance, it should work for global fleet tracking providers who use global SIMs from IoT providers, and offer them to different businesses, such as cargo ships or local delivery vehicles. Our approach must customize deployment for groups of devices with similar traffic patterns, specifically focusing on mobility and traffic volume, without any prior knowledge of their specific applications. 

This is a significant challenge, since it requires an unsupervised approach to anomaly detection, and one that preserve the privacy of the IoT application. 
To achieve this, ANCHOR clusters devices with similar traffic patterns, and explores various configurations for training the anomaly detection solution globally, per cluster, or based on the cluster with the majority of devices.

\vspace{.5mm}
\noindent \textbf{Ensuring data fidelity and validity.} involves accurately representing IoT device service anomalies (fidelity) and effectively explaining IoT device connectivity (validity) using our data.
A significant amount of our effort goes towards understanding how best to transform our raw \emph{mobility and authentication signaling} data into a representation that can exploit the strengths of machine learning for anomaly detection. 
With ANCHOR, we explore two different data representation approaches, which allows us to strike the balance between the messy and volatile live data, and the structured and refined dataset we use for training.


\vspace{.5mm}
\noindent \textbf{Operational feasibility.}
The ANCHOR solution must bring additional value on top of the existing monitoring approaches the IoT provider already uses.
Moreover, our design must identify both known and never-before-seen incidents/anomalies, as long as the latter surface within the signaling dataset.
Though we built our validation ground-truth dataset using incidents that were reported in the ticketing system, this does not guarantee that we capture the exhaustive list of anomalies that the IoT devices can suffer. 
For testing our ANCHOR detection engine, we take a two-sided approach: (i) we validate high-confidence samples of anomalies (from incident ticketing information), and (ii) test directly with the engineering team to validate whether the anomalies ANCHOR detected proved to be harmful to the underlying MNO provisioning the communication SIM to the IoT device. 

In parallel, we must ensure seamless integration of ANCHOR with the provider’s existing monitoring and analytics infrastructure. This process also involves determining the optimal timescales for running the engine (e.g., hourly, daily). Given that data is collected every five minutes, the anomaly detection engine could theoretically operate at that frequency. However, since incidents affecting a large number of connected devices are already captured by the existing reactive rule-based alarms, the operations team has advised that running ANCHOR once a day is sufficient.


\vspace{.5mm}
\noindent \textbf{Usability of the output.} 
We must enable the operations team to build a trust relationship with the ANCHOR detection engine, which is of paramount importance when attempting to integrate such tools within operational systems. 
It is thus critical to provide a tool to the operations teams that identifies unknown incidents, and allows them to investigate potential causes. 
We evaluate the trade-off between using powerful (and somewhat obscure) deep learning approaches, and using ensemble methods (such as isolation forest) that offer more explainability. 
For this, we put effort into translating the vast expert operator knowledge into features engineered for ANCHOR (see Section~\ref{sec:features-branch}).
For instance, devices experiencing connectivity anomalies typically generate a high volume of control traffic related to authentication requests. This insight led us to develop features that count specific protocol messages. 

%% file: sections/04_model_design.tex
\vspace{-3mm}
\section{ANCHOR Pipeline Design}
\label{sec:model_design}

In this section, we expose our experience with the design of the ANCHOR pipeline, and discuss our choices in terms of data representation and models that tackle the requirements and challenges of adapting to a live production system. 

\begin{figure*}
    \centering
    \includegraphics[width=.8\linewidth]{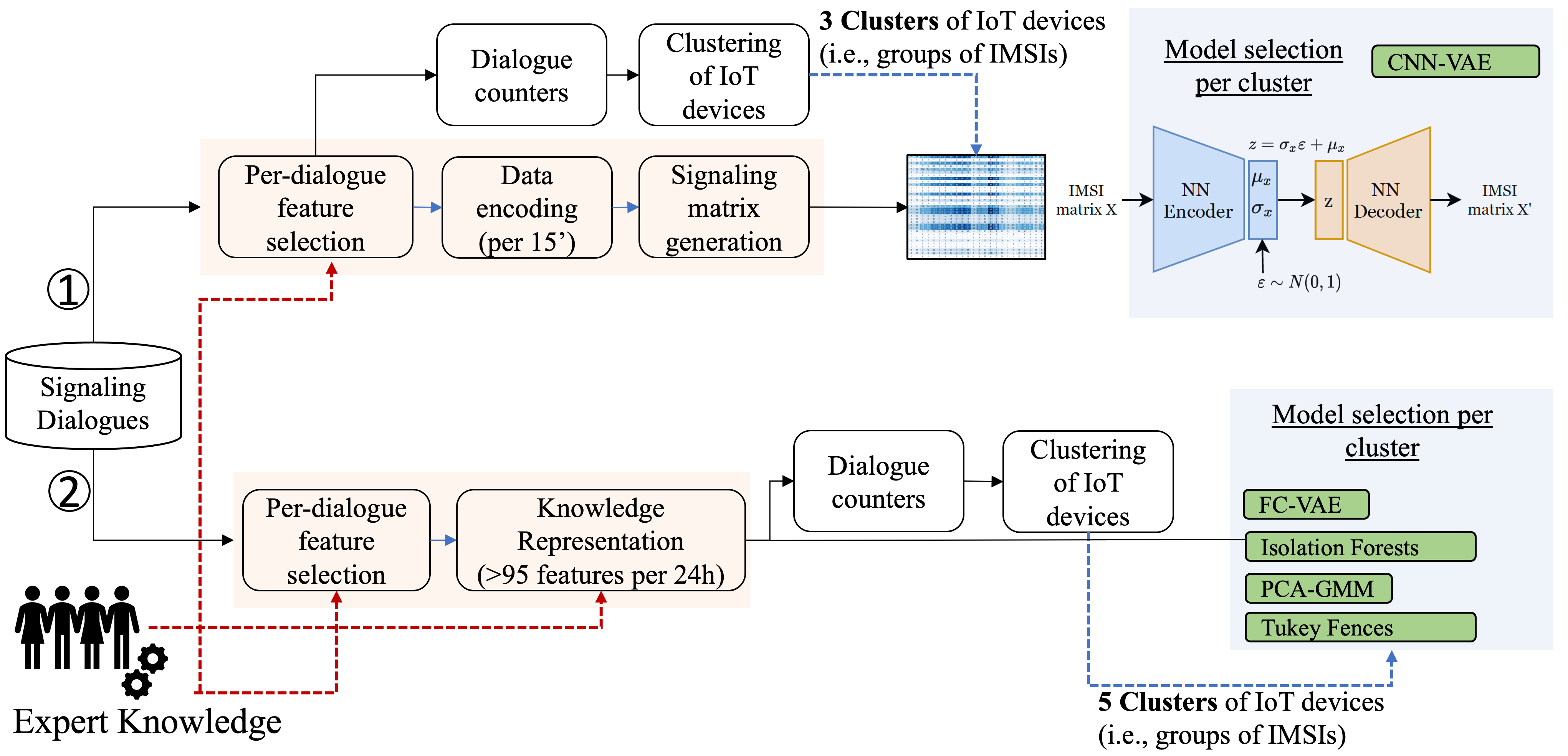} 
    \caption{\small ANCHOR Training Pipeline. Red lines represent information input by experts to better process the raw data, and the blue lines are context (clustering) extracted from the input features and passed to the DL/ML algorithms.
    } 
    \label{fig:framework}
    \vspace{-5mm}
\end{figure*}

With ANCHOR, we propose an unsupervised approach for anomaly detection of IoT devices, agnostic to device usage. 
The key idea of our solution is to learn the normal network behavior by estimating a low-dimension representation of the signaling messages exchanged by IoT devices.
Our methodology consists of three major steps (Figure~\ref{fig:framework}): dataset collection (i.e., the raw data in the form of signaling dialogues), data transformation (including feature selection and data representation, within the orange blocks in Figure~\ref{fig:framework}), and model selection (in the blue blocks in Figure~\ref{fig:framework}).  

We present next our experience with two different approaches for data representation, and designing the ANCHOR pipeline in consequence.

\vspace{-3mm}
\subsection{Signaling Matrix Representation}
\label{sec:vae-branch}

\begin{figure}[!t]
    \centering
    \includegraphics[width=\linewidth]{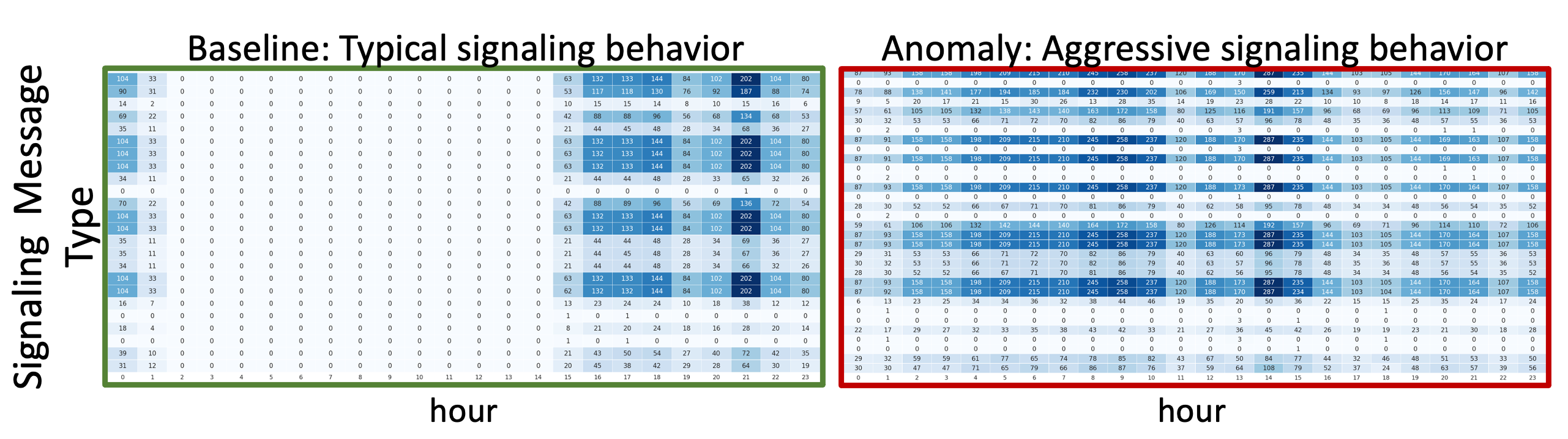}
        \vspace{-5mm}
    \caption{\small Example of aggressive signaling anomaly in an IoT device. Each tile in the image represents a signaling message counter (row) per 15' (column); we show a baseline day with typical signaling behavior (left image, green frame) vs. an anomalous day (right image, red frame).  }
	\label{fig:anom_example}
    \vspace{-5mm}
\end{figure}

In branch \circled{1} of the ANCHOR pipeline design (see Figure~\ref{fig:framework}), we build on the idea that deep learning’s success is mainly due to its ability to learn good representations from complex unstructured data, such as our IoT signaling datasets.
The lack of an exhaustive list of possible anomalies -- and of how these anomalies manifest in the control plane -- precludes us from attempting a supervised learning approach.
Based on the experts' input that some harmful anomalies appear as aggressive signaling, we focus on signaling volume, and encode the signaling events occurrences in a representation similar to images, which are generally suitable for deep learning models.
Figure~\ref{fig:anom_example} shows an example of aggressive signaling anomaly in the matrix representation for a single device. 

In the signaling matrices, the rows represent different protocol-specific features (e.g., the MAP and Diameter request types), and the columns represent time intervals (15 minutes, in our case). For example, following this logic, we group the values of "dialogue duration" by defining ranges that differentiate between normal and unusual values. 
In this case, a duration larger than 2,500 ms would be considered unusual, while being above 6 seconds would be rare. 
For each level we encode (i.e., add as a row in the signaling matrix), we then generate counters for presence within the time intervals we encode into each column of the signaling matrix. 
As the main objective for anomaly detection is intervention on the anomalous devices -- and such interventions within 24h are compliant with the needs of the engineering teams -- we focus on generating alarms for anomalies with daily resolution. 
However, we work with 15 minute intervals for encoding our data in columns of our 24h matrix, which allows us to gain information from the fluctuations during the day (Section \ref{sec:dataset}). 

\vspace{1mm}
\noindent\textbf{Clustering.}
We noticed that sub-groups of devices generate different signaling volumes (see Section~\ref{sec:dataset}), and propose an unsupervised clustering approach for uncovering the signaling patterns within our dataset. This step is crucial for obtaining a good performance model: if we trained the model with all devices, it would detect the minority class of devices as anomalies because of its different patterns.

The clustering we integrate on branch \circled{1} of our ANCHOR framework considers volume-related features for the devices we monitor. We use the number of MAP and Diameter messages per device and day for clustering. These features enable to distinguish between high or low signaling devices, and 2G/3G or LTE connected devices, which gives us enough information to divide them in three different kind of behaviors/clusters. Our clustering model is a GMM with as many components as days in the training dataset (i.e., 30 components if we use a month of data). We choose the number of clusters (namely, three clusters) based on their homogeneity and volume\cite{Kneedle, schwarz1978, akaike1974}. 
We generate the clusters using the training data. At inference, we use the previous day of data to assign devices to the generated clusters using the nearest neighbor algorithm\cite{cover1967nearest}, which classifies new device based on how similar devices (its neighbors) are classified.

\vspace{1mm}
\noindent\textbf{Anomaly Detection Models.}
\label{sec:vae-models}
Our anomaly detection approach builds on \ac{VAE}~\cite{kingma2013auto, rezende2014stochastic}. 
\ac{VAE} consists of two sub-networks, namely an encoder and a decoder. This architecture uses a bottleneck to learn a low-dimensional representation space $z$ of the training data $x$. The encoder learns the mapping  of the input data $x$ into a prior distribution $p(z)$, which is usually a standard Gaussian ($z \sim N(0,I)$). The last layer of the encoder outputs the parameters  $\mu_x, \sigma_x $ of the distribution $q(z|x)$. The decoder network is used to reconstruct $\hat{x} \sim p(x|z)$ from the latent representation $z$ sampled from $N(\mu_x,\sigma_x)$. 

In order to train a VAE, we need two losses; one is KL divergence loss $\mathcal{L}_{kl}$ , which is used to make sure that the latent vector $z$ is an independent unit Gaussian random variable, and the other is the reconstruction loss $\mathcal{L}_{rec}$, which measures how accurately the network reconstructed the input matrices. These two losses are combined as follows: $\mathcal{L} = \beta \cdot \mathcal{L}_{kl} + \mathcal{L}_{rec}$. There is a trade-off between how accurate our network can be and how close its latent variables can match the unit Gaussian distribution.  Higher $\beta$ gives more structured latent space at the cost of poorer reconstruction, while lower values give better reconstruction with less structured latent space. We set $\beta=1$, i.e., equal weight to both loss functions. 

We tested three different architectures, namely \ac{VAE} with Fully Connected neural networks, VAE with \acp{CNN}, and VAE with Recurrent Neural Networks, and selected the CNN-VAE as the best fit for the data signaling matrix representation, based on early testing results.
The \textbf{CNN-VAE} architecture we use includes four convolutional layers with different kernel sizes((25, 3), (3, 12), (5, 5), (3, 3)), and a stride fixed to 1, in both encoder and decoder networks. 
We use vertical filters to learn the relation between fields at same timestamps, while horizontal filters are used to learn the time-variant features.
Each convolutional layer is followed by a batch normalization layer, to help stabilize training and a ReLU activation layer~\cite{ioffe2015batch}. 
In the encoder, the CNN layers are followed by two fully-connected output layers (for mean and variance), used to compute the KL divergence loss\cite{kullback1951} and sample latent variable z. In the decoder, for up-sampling we apply a 2D transposed convolution operator with padding on the encoded data.



\vspace{-3mm}
\subsection{Expert Knowledge Representation}
\label{sec:features-branch}

In branch \circled{2} of the ANCHOR pipeline, we leverage the real-world expertise of the engineering team operating the IoT platform, and focus on transforming the expert knowledge into features for ANCHOR (see Figure~\ref{fig:framework}).
Various ongoing efforts are aimed at highlighting the importance of treating data as a first-class citizen in the development of advanced deep-learning solutions by ML/AI practitioners~\cite{sambasivan2021everyone, DR:HotNets-22}.
Here, we choose to focus on the data representation capability of capturing a wider set of potential anomalies (i.e., we design features that capture anomalies related to interactions between home and visited networks).


At inference stage, we compute an anomaly score for each of the devices by comparing new representations to the learned one. This offers the opportunity to detect anomalies caused by different unknown reasons, indistinct to the device purpose and without requiring any labeled data.

\vspace{1mm}
\noindent \textbf{Feature selection.}
We first transform the raw data (Section~\ref{sec:dataset}) into features optimized for clustering and anomaly detection. Leveraging insights from the IoT Provider’s operations team, we focus on protocol fields that enhance anomaly detection, including dialogue start time, duration, format violation indicators, request type (e.g., \ac{SAIR}, \ac{UL}), result code (success or rejection), and identifiers for home and visited operators, and core network elements involved.

In total, we include 95 features encompassing i) signaling traffic volume, ii) message types and signaling patterns, iii) device activity, iv) mobility statistics, and v) longitudinal activity statistics. The features cover different levels and directions to describe devices as holistically as possible from the perspective of the monitored signaling protocols.  

For each feature, we analyze the timeseries of daily records over a one-month period to compute statistical metrics. This approach aims to characterize overall device behavior and mitigate the impact of short-term anomalies. A complete list of features is available on GitHub.\footnote{https://github.com/phantom-owls/ANCHOR-features} 

\noindent\textbf{Clustering.}

As in branch \circled{1}, clustering is also essential in branch \circled{2} due to the diverse nature of signaling data. 
Since, for instance, combining logs from numerous sensors (with limited signaling volume and many stationary devices) with data from a few connected cars (which generate high signaling traffic due to mobility) would cause a global anomaly detection model to mistakenly identify connected cars as anomalies. 
In branch \circled{2}, rather than focusing solely on message volume, we incorporate a broader set of features. We use overall statistics --averaged over a one-month period-- on daily signaling dialogues, protocol usage, and mobility patterns for clustering. These features help differentiate between devices based on signaling volume, connection types (2G/3G vs. 4G LTE), and mobility status (stationary vs. mobile). This approach allows us to create clusters that better reflect the context of different IoT devices.

We generate clusters with Gaussian Mixture Model (GMM) configured to have as many components as days in the dataset (i.e., 30 components for a month of data). We choose the number of clusters based on homogeneity and volume, avoiding clusters with few devices; we determine this using the Bayesian Information Criterion (BIC).

\vspace{1mm}
\noindent \textbf{Anomaly Detection Models.}
\label{sec:models}
We use two different unsupervised models on our data: Isolation Forest, and \ac{VAE}. 
As baselines, we use Tukey Fences~\cite{tukey1977exploratory}, and \ac{GMM}.

The idea behind the \textbf{Isolation Forest~\cite{liu2012isolation}} algorithm is to isolate the anomalies in a dataset by creating a random forest~\cite{breiman2001random} of isolation trees. An isolation tree is a binary tree that recursively splits the nodes into smaller subsets. The splitting process is based on a randomly selected feature and a randomly selected split value for that feature. The path length from the root node to the terminal node where an instance is isolated is used to calculate an anomaly score. The anomaly score is a number between 0 and 1, and it represents the degree of anomaly of an instance. The closer the score is to 1, the more anomalous the instance is considered to be. The average path length of the trees in the forest is used to determine the threshold value for anomaly detection. The most important parameter in Isolation Forest is called \textit{contamination}, which represents the fraction of the dataset expected to contain anomalies. 
In ANCHOR, we set this value to $0.05$ (i.e., 5\%). This is of course a tunnable parameter: we set it up to 5\%, based on the operators' experience, who consider this percentage large enough to capture the majority of daily anomalies. We experimented with other contamination levels, such as 10\%, but found the results to be less accurate. 

We also implement \textbf{FC-VAE}, in which both encoder and decoder networks are composed of 4 fully connected layers with ReLU activation functions. The encoder reduces the number of features up to z=$32$ ($91 \times 128 \times 64 \times 32$) while the decoder does the reverse process, increasing the number of features at each layer ($32 \times 64 \times 128 \times 91$). To compare with the Isolation Forest model, we rank anomaly scores and classify the top 5\% as potentially anomalous devices.

Finally, for baseline we use a \textbf{PCA-GMM} implementation, where we first transform the features into a lower-dimensional representation using PCA, with the aim of retaining as much of the original variance as possible. We then cluster the reduced data using GMM, with the number of components in the mixture determined by the Bayesian Information Criterion (BIC).  In line with the other models, we rank each device by its probability score and select the bottom 5\%—those least likely to fit the mixture model—as potential anomalies. 
For thoroughness, we also use as baseline \textbf{Tukey Fences} -- also called the box plot rule~\cite{tukey1977exploratory} -- which rely on a simple rule: any data point beyond two statistical values ("fences") is considered an anomalous point. 

%% file: sections/05_ml_evaluation2.tex

\section{Evaluation Results}
\label{sec:ml_results}

In this section, we report on the usage of the anomaly solution in the production environment of the IoT provider. 
Together with the engineering team managing the IoT provider platform, we deployed ANCHOR in their system and validated the generated anomalies. 
Specifically, we executed a proof-of-concept for ANCHOR in October 2022, which we discuss next. To validate and understand the performance of each model on each cluster, we report the percentage of anomalies that were accurately detected from the ground truth per client/country/day.

%

\begin{table*}[!t]
    \footnotesize
        \caption{\small Percentage of anomalous devices classified as anomalous and present in tickets (i.e., recall), reported per IoT vertical client and day of validation in October. To indicate the order of magnitude of anomalies: * denotes >10 anomalies, \(\dagger\) denotes >100, and \(\ddagger\) denotes >1000 anomalies. Cell color indicates whether the detected anomalies exceed a pre-defined z-score threshold: blue for true positives (client with reported connectivity issues), red for false positives, and no color when the threshold was not surpassed.} 
    \label{tab:validation-branch3}
    \vspace{-4mm}
\begin{tabular}{p{3cm}|ll|l|lll|l|l|l}

 & \multicolumn{2}{l|}{\textbf{Client\#1}} & \multicolumn{1}{l|}{\textbf{Client\#2}} & \multicolumn{3}{l|}{\textbf{Client\#3}}  &  \multicolumn{1}{l|}{\textbf{Client\#4}} & \multicolumn{1}{l|}{\textbf{Client\#5}} & \multicolumn{1}{l}{\textbf{Control}} \\ 
\hline
                \textit{Date (dd/mm)} & \textit{5/10*} & \textit{6/10*}     & \textit{20/10*} & \textit{26/10\ensuremath{\ddagger}} & \textit{27/10\ensuremath{\ddagger}} & \textit{28/10\ensuremath{\ddagger}} & \textit{24/10*}        & \textit{10/10\ensuremath{\ddagger}} & \textit{20/10\ensuremath{\ddagger}}     \\ \cline{2-10} 
\hline
  \multicolumn{10}{c}{\textbf{\textit{Expert Knowledge - branch \circled{2}}} } \\\hline
{\textbf{IF global}}  &  0.0\% & 0.0\%  &  0.0\%  & 0.025\% & 0.069\%   & 0.0\%  & 0.0\%  &  \cellcolor{high}{8.51\%}  & 0.0\%     \\


{\textbf{IF per cluster}} & 6.52\%  & 6.25\% & 13.04\%  & \cellcolor{high}{8.24\%} & \cellcolor{high}{6.44\%}  & 4.34\%  & 1.54\% &  \cellcolor{high}{34.15\%} & 0.0\%            \\


{\textbf{IF major cluster}}   & 10.86\%  & \cellcolor{high}{25\%}  &  \cellcolor{high}{30.43\%}    & \cellcolor{high}{17.72\%}  & \cellcolor{high}{9.28\%}   & \cellcolor{high}{6.23\%}  & 6.79\% & \cellcolor{high}{8.51\%}  & 0.0\%   \\ 
\hline
{\textbf{GMM global}}  &  \cellcolor{high}{25.08\%} & 12.09\%  &  26.08\%  & \cellcolor{high}{20.87\%} & \cellcolor{high}{19.83\%}   & \cellcolor{high}{17.99\%}  & 0.61\%  &  1.84\%  & 0.51\%     \\

{\textbf{GMM per cluster}}  &  2.17\% & 6.25\%  &  4.34\%  & 1.73\% & 1.80\%   & 5.18\%  & 0.08\%  &  \cellcolor{high}{66.07\%} & \cellcolor{low}71.79\%     \\

{\textbf{GMM major cluster}}  &  2.17\% & 6.25\%  &  0.0\%  & 0.33\% & 0.48\%   & 4.20\%  & 0.0\%  &  \cellcolor{high}{64.51\%}  & \cellcolor{low}{65.23\%}     \\
\hline
{\textbf{FC-VAE global}}  &  19.57\% & 0.0\% & 0.0\% & 0.0\% & \cellcolor{high}{18.83\%} & \cellcolor{high}{18.06\%} & 0.0\% &  \cellcolor{high}{64.72\%}  & \cellcolor{low}65.54\%     \\
{\textbf{FC-VAE per cluster}}  &  17.39\% & 20.83\% & \cellcolor{high}{56.52\%} & 9.21\% & \cellcolor{high}{31.39\%} & \cellcolor{high}{28.09\%} & 0.0\% &   2.77\% & 2.77\%     \\
{\textbf{FC-VAE major cluster}}  & 0.0\% & 0.0\% & 17.39\% & 0.1\% & \cellcolor{high}{19.02\%} & \cellcolor{high}{13.6\%} & 0.0\% &  0.41\% & 0.0\%  \\
\hline
\multicolumn{10}{c}{\textbf{\textit{Raw counters - branch \circled{1}} }} \\\hline
\cline{2-10} 
\hline
{\textbf{CNN-VAE global model}}  &  17.64\% & 0.0\% & 13.04\% & 9.83\% & 9.84\% & 10.25\% & 15.38\% &  10.00\% & 10.21\%     \\
{\textbf{CNN-VAE per cluster}}  &  2.94\% & 11.11\% & 13.04\% & 10.35\% & 9.60\% & 9.90\% & 2.19\% &  2.66\% & \cellcolor{low}{17.77\%}     \\
{\textbf{CNN-VAE major cluster}}  &  2.94\% & 0.0\% & 21.73\% & 10.35\% & 9.58\% & 9.93\% & 10.98\% &  9.66\% & 10.49\%     \\
\hline
\end{tabular}
\end{table*}

\begin{table}[!t]
\vspace{-2mm}
	\footnotesize
  \caption{\small Mean values for features that distinguish clusters on branch 2 of ANCHOR.}
  \label{tab:overview_clustering}
  \vspace{-2mm}
  \centering
  \begin{tabular}{|p{2.9cm}|r|r|r|r|r|}
  
    \hline
  Device Cluster & 1 & 2 & 3 & 4 & 5 \\ 
    \hline
    \hline
  \rowcolor{shadecolor}\multicolumn{6}{c}{Overall Statistics: Protocol Usage and Devices} \\ 
    \hline
\# Devices & 1.7M & 420k & 894k & 300k  & 621k \\
\# Dialogs. per device/day & 8.7 & 22.4 & 98.4 & 1300& 72\\
\ac{MAP} Dialogs. (2G/3G) [\%] & 100 &  0 & 99.99 & 63.4 & 18.1 \\ 
Diameter Dialogs (LTE) [\%] & 0 & 100 & 0.01  & 36.6 & 81.9 \\ 
    \hline

   \rowcolor{shadecolor}\multicolumn{6}{c}{Dialogue Type Volume} \\ 
    \hline 
  \#\ac{SAI} (2G/3G) & 4.2 & 0 & 46.4 & 306 & 2.1 \\
  \#\ac{SAI} (LTE) & 0 & 15 & 0.01 & 291 & 43 \\
  \#\ac{UL} (2G/3G) & 1.2 & 0 & 18,5 & 145.5 & 1.1 \\
  \#\ac{UL} (LTE) & 0 & 4.4 & 0.001 & 244 & 12.7  \\
  \#\ac{CL} (2G/3G) & 1.7 & 0 & 21.6 & 117.1 & 1.6 \\  
  \#\ac{CL} (LTE) & 0 & 1.2 & 0.001 & 102.1 & 6.5 \\
    \hline

  \rowcolor{shadecolor}\multicolumn{6}{c}{Mobility Features} \\ 
    \hline
LTE Operator Changes & 0 & 0 & 0 & 25 & 1.6\\ 
\ac{VLR} Changes & 2.2 & 0 & 36.3 & 278 & 1.12 \\
\ac{SGSN} Changes & 1.51 & 0 & 20.5 & 103.8 & 1 \\
    \hline

  \end{tabular}
  \vspace{-5mm}
  \end{table}


  
\noindent\textbf{General Clusters.} 
During September 2022, the clustering algorithm segmented devices from various customers into distinct groups, depending on the chosen branch (see Figure \ref{fig:framework}). Next, we describe five clusters identified in branch \circled{2}, which uses features that define device context, mainly based on communication frequency and mobility. Table~\ref{tab:overview_clustering} provides an overview of these five clusters and their main characteristics. The following patterns emerge:

\noindent\textit{Cluster 1}: predominantly stationary devices, 2G/3G connectivity only, low volume of dialogues

\noindent\textit{Cluster 2}: stationary devices, 4G LTE connectivity only, medium volume of dialogues

\noindent \textit{Cluster 3}: mildly mobile devices, 2G/3G connectivity, high volume of dialogues

\noindent \textit{Cluster 4}: highly mobile devices, mixed 2G/3G/4G LTE connectivity, very high volume of dialogues

\noindent \textit{Cluster 5}: stationary devices, patchy 4G LTE connectivity (with 2G/3G), high volume of dialogues

We validated the consistency of the clustering from Table~\ref{tab:overview_clustering} between two consecutive months (September and October 2022).
For the studied clients, presented in table \ref{tab:validation-branch3} we found that for most clients the IoT devices with anomalies map to clusters \textit{c2}, \textit{c4} and \textit{c5}, with the vast majority (in the order of thousands of devices) falling in \textit{c2} (i.e., devices that are mostly stationary, with 4G LTE connectivity only, and medium volume of dialogues). For cluster \textit{c4}, we have the lowest number of anomalous devices mapped, with as little as two IoT devices in Spain or Saudi Arabia. The distribution is similar for cluster \textit{c5}, with the exception of India, where the number of IoT devices with ground-truth anomalies is roughly half of those in \textit{c2} (in the order of thousands).

\noindent\textbf{Training and Inference.}
Training time for isolation forest varies between 50 and 400 minutes. However, for the VAE model, the training time could be up to 2 days for creating a general model without omitting the clustering step (global model). 
Same as before, we leverage for this task a processing instance with 56 cores Intel Xeon E7, 504GB of RAM, and 7 GPUs NVIDIA GeForce RTX 2080.
The inference time has small variability and is usually well below the upper bound of a few tens of minutes.
In both inference and training time, we do not consider the time it takes to create the features, since this is done automatically by a batch job, and does not represent any overhead. 
Nevertheless, we do consider the time to transform the metrics (for Isolation Forest and VAE, we use the $StandardScaler$ transformation), as well as the time to merge the dataset output by the clustering algorithm and the features for the different ML models. 

\definecolor{blueish}{rgb}{0.9, 0.9, 0.99} 
\definecolor{redish}{rgb}{0.98, 0.91, 0.9} 

\noindent\textbf{Statistical significance.} The clients we evaluated had nearly all their devices behaving anomalously on specific days, and in specific deployments per country. Detecting this type of anomaly is crucial, as it can jeopardize the contract between the client and the IoT connectivity provider. However, the incidents we consider in table~\ref{tab:validation-branch3} did not trigger any alarms in the current system. 
This is understandable given the large number of devices (4 million), where only a few tens, or at most a few thousand, may behave anomalously for a given client. 
The challenge is compounded by the heterogeneity of clients. While we do not expect to achieve a recall close to 100\%, our goal is to provide a statistically significant signal to assist operators in identifying and addressing these anomalies. 
To achieve this, we set up a baseline by comparing against a uniform distribution, where anomalies are evenly distributed among clients. 
Since the number of potential anomalies is a tunable parameter (e.g., the contamination factor in Isolation Forest), we compare whether the number of detected anomalies is statistically significantly larger than those from a uniform distribution.

Specifically, we calculated the z-score for each client to measure the difference between the number of expected anomalous devices and the number of devices labeled as anomalous by the predictive model: \(
Z_i = \frac{D - E}{\sqrt{E \times Ci}}\), where \(E\) is the number of expected anomalous devices, \(D\) is the number of devices labeled as anomalous, and \(Ci\) is the confidence interval. 
For each client, we compared the data against the 99\% confidence interval.\footnote{This is a parameter that can be adjusted by the operator} 
For example, if a client has an expected number of 50 anomalous devices (as per uniform distribution) and our model labels 80 devices as anomalous, the z-score would be \(\pm 4.24\), which is greater than \(\pm 2.576\) (the z-score for a 99\% confidence interval). Since the z-score exceeds the \(\pm 2.576\) threshold, it suggests that the observed anomalies are not due to random chance but is likely a genuine issue. 
In table \ref{tab:validation-branch3}, we highlight the cases where the z-score for a given client exceeds the 99th percentile threshold. 
We highlight in \sethlcolor{blueish} \hl{light blue} instances when this occurs for a client with known anomalies. Conversely, we highlight in \sethlcolor{redish} \hl{light red} instances when this occurs for the control client -- a client that was confirmed by the operational team to have behaved as expected during that day -- indicating a false positive.

\noindent\textbf{Effect of clustering.} 
We aimed to evaluate the importance of the clustering step by comparing its performance with a \textit{global} model that excludes clustering and uses all the data in a single model. In contrast, the \textit{per cluster} model applies the specific model associated with each device's assigned cluster.
Nevertheless, while the clustering step aims to optimize model selection, it may not be accurate, and some devices might fall on the border between two clusters. To address this, we evaluated a model that employs the model of the cluster where the majority (\textit{major cluster}) of devices from a given client in a specific country belong. For example, if Client\#1 in Spain has 1000 devices mapped to cluster \textit{c1}, 70 devices mapped to \textit{c2}, and no devices mapped to any other cluster, we exclusively use the model of \textit{c1}, for detecting anomalies for this client in that specific country. 

\noindent\textbf{Branch \circled{2} outperforms branch \circled{1}}. Results from the expert knowledge representation approaches in Branch \circled{2} are superior, enabling models to detect IoT clients with anomalous devices more effectively than those using raw counter data in Branch \circled{1}. For Branch \circled{1}, we primarily focused on CNN models and observed that none of the model variations detected enough anomalies to surpass the 99th percentile z-score threshold. Worse, one model incorrectly flagged the control client (verified by the operator to have no anomalies) as anomalous.

\noindent\textbf{Isolation Forest (IF) results.} 
Overall, the Isolation Forest model demonstrates superior performance, particularly in cluster \textit{c2}, consistently triggering alarms for a significant percentage of affected IoT devices, as indicated by the ground truth dataset. In specific regions, such as the United States and Argentina, Isolation Forest successfully triggers alarms for over 60\% of devices for Client\#2 and 30\% for Client\#5. This high alarm rate was later validated to be statistically significant. Importantly, the operations team can choose how to respond to ANCHOR's results. We have set a conservative threshold (99th percentile z-score) to minimize false positives and reduce the burden on operational engineers. Even with this conservative threshold, Isolation Forest detects anomalous behavior in 4 out of 5 clients at best, and in 2 out of 5 clients at worst, without flagging the control client as anomalous (i.e., no false positives). Client \#4 was particularly challenging, with fewer than 100 devices behaving anomalously. The poorest performance for Isolation Forest occurs when using a global model without the clustering step, identifying only 1 out of 5 anomalous clients. This underscores the critical role of clustering. 

\noindent\textbf{Probabilistic Models.} GMM and FC-VAE produced worse results compared to IF. Surprisingly, GMM outperforms Isolation Forest and FC-VAE for certain clients, particularly when dealing with a limited number of ground-truth anomalies (less than 100). We conjecture this is due to the stochastic nature of the data. Additionally, for FC-VAE, the benefit of the clustering step is evident, as the global model is the only one among the three FC-VAE presented variants that incorrectly label the control client as anomalous. In contrast, IF consistently produces reliable results across different clients and dates.

%% file: sections/06_discussion.tex
\section{Discussion and Experience}
\label{sec:discussion}


Anomaly detection in a live production system is a challenging task. In this section, we discuss our experience with the design and testing ANCHOR, as well as expose some of its limitations and future steps. 

\vspace{-3mm}
\subsection{Live-test trial}
The last step once models are learned, is to use them to perform inference. 
We live-tested ANCHOR during February 25th 2023 using the isolation forest model, and analyzed the validity of the top 20 ranked devices as anomalous in cluster \textit{c2}. 
In addition to the previous analysis, where we validated that the models are at least able to detect known anomalies, here we aim to detect completely unknown anomalies. 
We validate together with the engineering team 16 out of the 20 top devices ANCHOR highlighted. 
Three of them belong to an IoT customer deploying devices in India, while the other 13 belong to a different client connecting devices over 7 different countries. 
Overall, we identify six of them as potentially harmful (three in each IoT customer). 
When engineers receive the anomaly alarms, they follow a structured process to manage and resolve the issue efficiently. First, they perform manual checks on specific devices, verifying the raw data and investigating potential causes. 
This approach has proven effective in numerous historical cases, where manual intervention allowed us to identify and address anomalies that might have otherwise gone unnoticed.
In addition to manual checks, the engineers adopt a proactive approach by reaching out to clients directly. By engaging with clients, they gain a deeper understanding of the nature of the anomaly reported by ANCHOR, ensuring a quicker and more accurate resolution.
For instance, for few devices with aggressive signaling behavior that were flooding the signaling platform, the engineers collaborated with the IoT client to solve the issue.  

It is worth noticing that, despite using in February 2023 a model trained in September 2022, it is still possible to identify additional harmful devices on the top of regular reactive mechanisms previously deployed by the operations team.
Classic ML challenges on how to keep the models up-to-date clearly apply here, but are out of scope in this paper.

\vspace{-3mm}
\subsection{Signaling Data Representation}

We discuss next our experience with exploring and testing alternative approaches for data representation to the ones we presented in Section~\ref{sec:model_design}. 
Specifically, we tested a new data representation designed for sequence-to-sequence (Seq2Seq) models~\cite{sutskever2014sequence} to take advantage of the temporal patterns within the data, and the sequential logic of the signaling protocols we monitor.


Seq2Seq models have become popular for applications such as speech-to-text \cite{bahdanau2016end}, text-to-speech, and neural machine translation~\cite{britz2017massive, wu2016google}. This model uses a \ac{LSTM} to map the input sequence to a vector of fixed dimensionality, and then another deep LSTM to decode the target sequence from the vector. The LSTM’s ability to successfully learn on data with long-range temporal dependencies makes it a natural choice for this application due to the time lag between samples.
The input to the Seq2Seq model are three different vectors: the encoded sequences of request types and result codes ($codes$ $x$); its corresponding timestamps $t$; and the difference between consecutive timestamps, which we call $deltas$. We encode deltas by using the same mapping we used for duration, and consider an extra category when the previous code is from a different day.



Using this data representation, we ran early testing using a limited dataset (i.e., anomalies collected over only two days). 
This approach yielded underwhelming results, capturing non-harmful anomalies at a very high training and inference cost.
Moreover, we found that such complex models lack explainability, making simpler approaches such as IF models more fitting for our case. 
We thus abandoned this effort. 

In comparison, we found the CNN-VAE is the model that is able to detect a higher number of anomalies over the same two-days dataset in this early testing phase.
However, given the matrix of raw-counters representation of our signaling information, most anomalies ANCHOR with CNN-VAE detected were related to aggressive signaling behavior from IoT devices. 
In fact, thanks to ANCHOR, we were able to detect (previously undetected) IoT devices that were stressing the visited networks to the extent of endangering the commercial roaming agreement these had with the home operator supporting the IoT provider. 
These devices were generating large volumes of signaling against local providers in the attempt to complete the attach procedure. 
Based on these results, the IoT provider operational team decided to integrate this version of the ANCHOR solution within its monitoring system, as an aggressive device flag. 
Despite its utility, this initial version of ANCHOR was limited to focus only on anomalies related to the volume of signaling IoT devices generate, justifying the choice of exploring the alternative of feature engineering.
 


\vspace{-3mm}
\subsection{ANCHOR Generalization}

ANCHOR leverages only control-plane information corresponding to the devices the IoT Provider we study connects. 
We collect our dataset from the roaming hub that interconnects the IoT Provider with other partner \acp{MNO} globally. 
Roaming hubs bring the potential to serve \ac{IoT} verticals in every location, bridging the gap to seamless roaming (Section~\ref{sec:background}).
Any commercial IoT Provider uses at least one roaming hub~\cite{lutu2020ipx, lutu2020things}, and thus use the same control plane signaling protocols we analyzed here to operate globally.
Recent work~\cite{geissler24untangling} has compared the signaling traffic patterns in the signaling traffic from two different commercial IoT Providers. 
After separately processing two independent, large-scale signaling traffic datasets, the authors reported that similar signaling characteristics are well-suited to identify groups of devices across both datasets. 
Moreover, their work also supports that domain expert knowledge is paramount in building a meaningful set of features, that capture the behavior of myriad IoT devices, even across different IoT provider systems.
This strongly supports the potential generalization of the ANCHOR solution: as long as we are able to extract similar datasets (see Section~\ref{sec:dataset}), we can undoubtedly apply the ANCHOR approach to any other IoT provider in the ecosystem. We leave this for future work.  






%% file: sections/07_related.tex
\section{Related Work}

When it comes to approaches for anomaly detection in literature, there are two main dimensions along which to distinguish existing research: the applied methodology and the underlying use case. 

In the area of statistical models, several approaches have been proposed in the past~\cite{fahim2019anomaly, chandola2009anomaly, kalinichenko2014methods, eltanbouly2020machine, wang2021machine}. Specifically, when not limiting to the IoT or mobile area, approaches like ANOVA, ARIMA, Gaussian Mixture Models (which are also used in this work), Bayesian Models, or Dissimilarity Measures~\cite{fahim2019anomaly} have well investigated in the past. More recently, studies focusing on network anomalies, and specifically anomalies in the area of IoT have appeared~\cite{cook2019anomaly, diro2021comprehensive, zhang2021time, chatterjee2022iot}. Those studies, and many others, typically aimed to reduce or eliminate the need for human interaction and expert knowledge in deploying mechanisms for specific use cases. In contrast, this work argues that expert input remains crucial, especially given the scale of managing millions of devices. Therefore, we advocate for a model that incorporates features designed by engineers, which not only provides insights but also allows experts to understand and guide the process of addressing anomalies effectively.

Furthermore, various advanced machine learning models have been explored, including Support Vector Machines~\cite{bellini2020anomaly}, k-means Clustering, k-Nearest Neighbors, Local Outlier Factor Models \cite{braei2020anomaly}. Additionally, deep learning models like Deep Convolutional Networks \cite{chalapathy2019deep}, Recurrent Neural Networks~\cite{zhang2021time}, \ac{LSTM} models~\cite{yang2021anomaly} and many others have been investigated in the past. These studies typically focus on user plane traffic, which involves payload-related communication of devices, and use datasets limited to hundreds or up to a few hundred thousand data points. In contrast, our work addresses anomalies in the signaling plane of mobile networks using a large-scale dataset of globally distributed devices. To our knowledge, this is the first study to analyze a large-scale signaling dataset for mobile roaming with a focus on unsupervised anomaly detection.




%% file: sections/08_conclusions.tex
\vspace{-3mm}
\section{Conclusions}
\label{sec:conclusions}

In this paper, we shared our effort on designing, implementing and deploying an anomaly detection pipeline within a live IoT provider, which connects millions of devices world-wide. 
The complex cellular ecosystem and its roaming function support the IoT provider, and the operations teams deploy various processes to ensure the availability of their service. 
Despite state-of-the-art alarm systems in place, anomalies do still occur in practice.
The goal of ANCHOR was the realization of a practical solution that is able to support current efforts of operations teams by detecting anomalies that usually escape alarm systems before they become critical (and then receive attention after the request of the clients). 
To achieve this, we showed that exploiting expert knowledge to engineer meaningful features that then enable operators to reason about the identified anomalies is a promising approach. 
ANCHOR implements device-level anomaly detection using signaling conversations between core network functions that enable global IoT device connectivity, which we process into representations that optimize machine learning techniques.
We created a solution that works on different radio technologies and protocols, heterogeneous IoT applications, and at large scale (billions of daily messages). We prioritize control plane data for lightweight monitoring, addressing cost constraints and privacy concerns.

ANCHOR relied on human operators in the “slow loop”, which was beneficial to separate harmful and non-harmful anomalies, and provide guidelines to improve the machine learning approaches. With this, it becomes apparent that the amount of time spent by human experts in the troubleshooting process is a valuable resource that anomaly detection systems should explicitly take into account~\cite{navarro2022human}. For the future, we plan to evolve ANCHOR to benefit from the direct feedback of human operators. Moreover, to successfully evolve ANCHOR beyond the approach we presented in this paper, we also plan to address the human operators' need for trustworthiness and explainability, as well as test ANCHOR for other IoT providers.